\documentclass[aps,prb,twocolumn, amsmath,amssymb,superscriptaddress]{revtex4-1}

\usepackage{amsmath}
\usepackage{amssymb}
\usepackage{amsthm}
\usepackage[pdftex]{color}
\usepackage{graphicx}% Include figure files
\usepackage{subfigure}
\usepackage{dcolumn} % Align table columns on decimal point
\usepackage{bm} % bold math
\usepackage{epic}
\usepackage{longtable}
\usepackage{ulem}   % to strike things out
\usepackage{hyperref}
\newcommand{\q}{\textbf{q}}
\newcommand{\veta}{\vec{\eta}}

\begin{document}
\title{
Fluctuation and strain effects in a chiral $p$-wave superconductor
}

\author{Mark H Fischer}
\author{Erez Berg}
\affiliation{Department of Condensed Matter Physics, Weizmann Institute of Science, Rehovot 7610001, Israel}

\begin{abstract}
For a tetragonal material, order parameters of $p_x$ and $p_y$ symmetry are related by rotation and hence have the same $T_{\rm c}$ at a mean-field level.
This degeneracy can be lifted by a symmetry-breaking field, like (uniaxial) in-plane strain, such that at $T_{\rm c}$, the order parameter is only of $p_x$ or $p_y$ symmetry. 
Only at a lower temperature also the respective other order parameter condenses to form a chiral $p$-wave state.
At the mean-field level, the derivative of $T_{\rm c}$ with strain is discontinuous at zero strain. 
We analyze consequences of (thermal) fluctuations on  
the strain-temperature phase diagram within a Ginzburg-Landau approach. 
We find that the order-parameter fluctuations can drive the transition to be weakly first order, rounding off this discontinuity.
We discuss the possibility of a second-order transition into a non-superconducting time-reversal-symmetry-breaking phase and consequences for the spin-triplet superconductor Sr$_2$RuO$_4$. 
\end{abstract}

\maketitle
\section{Introduction}
\begin{figure}[b]
  \centering
  \includegraphics{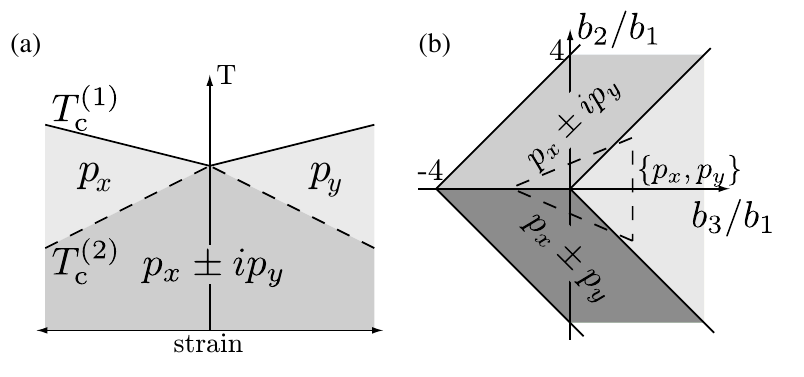}
  \caption{(a) Mean-field temperature-strain phase diagram for the weak-coupling parameters $b_2=-b_3=2/3b_1$ and (b) phase diagram for s=0 with respect to the fourth-order terms of the free energy. For parameters inside the dashed triangle, fluctuations cannot drive the superconducting transition.}
  \label{fig:params}
\end{figure}

In a tetragonal superconductor, order parameters of $p_{x}$ and $p_y$ symmetry are degenerate as they are related by a $C_4$ rotation. This allows for a time-reversal-symmetry (TRS) breaking order parameter of (chiral) $p_{x}\pm i p_{y}$ structure at $T_{\rm c}$,~\cite{sigrist:1991} as possibly realized in Sr$_2$RuO$_4$.~\cite{kallin:2012} When $C_4$ symmetry is broken, e.g., by in-plane strain, the degeneracy is lifted and the now distinct order parameters have different critical temperatures. Within a mean-field picture, this leads to a linear increase of $T_{\rm c}$ with strain and a cusp around zero strain, see Fig.~\ref{fig:params}(a). A recent strain study on Sr$_2$RuO$_{4}$ has %, 
indeed found a substantial enhancement of $T_{\rm c}$ for both tensile and compressive strain.~\cite{hicks:2014} 
However, no sign of a cusp of $T_{\rm c}$ around zero strain was observed, thus raising the question whether such behavior is consistent with a chiral $p$-wave superconductor.

Motivated by this experiment, we analyze how thermal order-parameter fluctuations can change the mean-field temperature-strain phase diagram. Fluctuations couple the order parameters of $p_{x}$ and $p_{y}$ symmetry and can thus both enhance or suppress the effect of strain. Moreover, as a chiral superconducting state not only breaks $U(1)$ symmetry, but also $Z_2$ (TRS), fluctuations associated to the latter possibly lead to a non-superconducting TRS-breaking phase at 
temperatures above the superconducting $T_{\rm c}$.~\cite{nandkishore:2012b, bauer:2013} Such behavior has also been found for multi-band superconductors with frustrated interband coupling~\cite{bojesen:2013, bojesen:2014} and is sometimes referred to as having a ``preemptive'' \cite{fernandes:2012b} or ``vestigial'' \cite{nie:2014} phase. Alternatively, the fluctuations can also drive the transition to be (weakly) first order. In that case, the strain has to overcome a finite strength before breaking up the chiral superconducting state and leading to a double transition, thus removing the cusp at zero strain.

In this work, we employ a variational analysis of the free energy based on Ginzburg-Landau theory describing a (strained) two-component order parameter. 
Analyzing the saddle-point (self-consistent-field) equations, we derive the conditions for TRS-breaking fluctuations to drive the system to superconductivity for small strain and discuss the possibility of a TRS-breaking, non-superconducting phase. A similar analysis has been performed for the pnictides, where the instability toward striped magnetic order is generally driven by a nematic instability.~\cite{Fang2008, Xu2008, fernandes:2012b} Finally, we present the full temperature-strain phase diagram and discuss the relevance of our results for Sr$_2$RuO$_4$.

We investigate a two-component order-parameter $\vec{d}(\vec{k}) = \hat{z}(\eta_x k_x +  \eta_y k_y)$ reflecting the symmetry possibly realized in Sr$_{2}$RuO$_{4}$. The Ginzburg-Landau-type free-energy density reads\cite{sigrist:1991}
\begin{equation}
  f[\vec{\eta}] = f_{2}[\vec{\eta}] + f_4[\vec{\eta}]+ f_{\rm grad}[\vec{\eta}], 
  \label{eq:genaction}
\end{equation}
with
\begin{eqnarray}
  f_{2}[\vec{\eta}] &=& a(|\veta|^2) + s(|\eta_x|^2 - |\eta_y|^2)\label{eq:GLF2},\\ 
  f_{4}[\veta] &=& b_1|\veta|^4 + \frac{b_2}{2}(\eta^{*2}_x\eta_y^2 + {\rm c.c.}) + b_3|\eta_{x}|^2|\eta_{y}|^2, \label{eq:GLF4}\\ 
  f_{\rm grad} &=& K_1(|\partial_x\eta_x|^2 + |\partial_y\eta_y|^2) + K_2(|\partial_y\eta_x|^2 + |\partial_x\eta_y|^2)\nonumber\\
  &&+ [K_3(\partial_x\eta_x)^*(\partial_y\eta_y) + K_4(\partial_y\eta_x)^*(\partial_x\eta_y) +\rm{c.c.}]\nonumber\\ 
  &&+ K_{5}(|\partial_z\eta_x|^2 + |\partial_z\eta_y|^2),
  \label{eq:GLFgrad}
\end{eqnarray}
where $a=a_0(T-T^{(0)}_{\rm c})$, $b_i$, and $K_i$ are phenomenological parameters. Note that the second term of $f_2[\veta]$ describes the coupling of the uniaxial strain $s$ along the $(1,0)$ crystalline axis to the order parameter.~\cite{sigrist:2002} Strain along the $(1,1)$ axis can be introduced through a coordinate transformation or a term $s' (\eta_x\eta_y^* + \eta_y\eta_x^*)$. For simplicity, we have absorbed 
the coupling constant between strain and the superconducting order parameter into $s$.

Figure~\ref{fig:params}(b) shows the mean-field phase diagram for the free energy density of Eq.~\eqref{eq:genaction} without applied strain. As we are interested in the chiral $\veta = \eta_0 (1, \pm i)$ solution, we consider only $b_2 > b_3$. Further, the stability condition for the fourth-order terms requires $4b_1 - b_2 + b_3 > 0$.
For finite strain, the transition splits and the system undergoes a first transition at $T^{(1)}_{\rm c}(s) = T^{(0)}_{\rm c} + |s/a_0|$. Only at a lower temperature 
\begin{equation}
  T^{(2)}_{\rm c}(s) = T^{(0)}_{\rm c} - \left| \frac{s}{a_0} \right| \frac{4b_1 - b_2 + b_3}{b_2 - b_3}
\end{equation}
the system enters a chiral phase, see Fig.~\ref{fig:params}(a). Note that for two completely decoupled order parameters, i.e., $b_3=-2b_1$ and $b_2=0$, $T_{\rm c}^{(2)}(s) = T^{(0)}_{\rm c} - |s/a_0|$.

\section{Method}
In the following, we use a self-consistent harmonic variational approach~\cite{[{See, e.g., }] giamarchi:2003} to find the phase diagram described by the free energy 
\begin{equation}
  F = \langle h \rangle - T S = T (\langle f\rangle - S),
  \label{eq:free0}
\end{equation}
with $h=T f$ the Hamiltonian density of the system, $S$ the entropy and $T$ the temperature. The expectation value $\langle.\rangle$ in Eq.~\eqref{eq:free0} is evaluated as
\begin{equation}
  \langle A \rangle = \frac{1}{Z} \int (D\veta) A \rho[\veta],
  \label{eq:expvalue}
\end{equation}
with  $\rho[\veta] = \exp(-f[\veta]/T$ ) the Boltzmann distribution function, $Z = \int (D\vec{\eta}) \rho[\vec{\eta}]$, and $A$ any functional of the fields $\veta$.
We approximate $\rho[\veta] \approx \rho_\Psi[\veta] = \exp(-f_{\Psi}[\veta]/T)$, where $f_{\Psi}[\veta] = f_2[\veta]+f_{\rm grad}[\veta] + \veta^\dag \underline{\Psi} \veta$ is a quadratic variational free energy with $\underline{\Psi}$ a variational $2\times2$ matrix. 
The free energy can thus be written as
\begin{eqnarray}
  F_{\Psi} &=& \langle f \rangle_{\Psi} - T S_\Psi\nonumber\\
  &=& \langle f_{\Psi} \rangle_{\Psi} - T S_{\Psi} + \langle f - f_{\Psi} \rangle_{\Psi},
  \label{eq:variationalF}
\end{eqnarray}
where $S_{\Psi}$ is the entropy corresponding to the quadratic action.
The first two terms on the right-hand side are simply $F^0_{\Psi}$, the free energy corresponding to $f_\Psi[\veta]$, while the last one can be decoupled, since it is evaluated over the Gaussian distribution function $\rho_\Psi[\veta]$. Finally, we minimize with respect to all fields. This approach allows us to approximate the free energy both in the normal and the ordered, i.e., superconducting state. In Appendix~\ref{app:hubb}, we will comment on its relation to another decoupling scheme, namely a Hubbard-Stratonovich approach in large $N$, the number of flavors of each field $\eta_{x, y}$ (with $N=1$ the physical value).

Concretely, starting from the disordered side, $\langle \veta\rangle =0$, $f_{\Psi}[\veta]$ reads in momentum space
\begin{equation}
  f_{\Psi}[\vec{\eta}] = \int (d^3q)\vec{\eta}^\dag \underline{G}_{\q}^{-1}\vec{\eta},
  \label{eq:quadratic}
\end{equation}
where $\underline{G}_{\q}^{-1} = [f_0(q)\underline{\tau}^0 + \vec{f}(q)\cdot\vec{\underline{\tau}}]$, $\underline{\tau}^0$ and $\vec{\underline{\tau}}$ are the $2\times2$ identity and Pauli matrices, respectively, and we used the short form $(d^3q)=d^3q/(2\pi)^3$. Further, 
\begin{eqnarray}
  f_0(\q) &=& a + \psi_0 + \frac{K_1+K_2}2 (q_x^2 + q_y^2) + K_5 q_z^2, \\
  f_1(\q) &=& \psi_1 +  (K_3+K_4)q_xq_y,\\
  f_2(\q) &=& \psi_2,\\ 
  f_3(\q) &=& s + \psi_3 + \frac{K_1-K_2}2 (q_x^2 - q_y^2),
  \label{eq:fs}
\end{eqnarray}
with $\underline{\Psi}=\psi_0\underline{\tau}^0 + \vec{\psi}\cdot\vec{\underline{\tau}}$.
It therefore follows that
\begin{equation}
  F^0_{\Psi} = T \int (d^3q) \log[f_0(\q)^2 - |\vec{f}(\q)|^2]
  \label{eq:varF}
\end{equation}
and using
\begin{equation}
  \langle\vec{\eta}\vec{\eta}^\dag\rangle_{\Psi} = \int (d^3q) \underline{G}_q = g_0\underline{\tau}^0 + \vec{g}\cdot\vec{\underline{\tau}},
\end{equation}
we can factorize the fourth order terms to find
\begin{eqnarray}
  \frac{F_\Psi}{T} &=& \int (d^3q) \log[f_0(\q)^2 - |\vec{f}(\q)|^2]\nonumber\\
   &&+ (6b_1 + b_3) (g_0)^2 - 2 \psi_0g_0\nonumber\\
   && + (2b_2 + b_3 + 2b_1) (g_1)^2 - 2 \psi_1 g_1\nonumber\\
   && + (b_3 + 2 b_1 - 2b_2) (g_2)^2 - 2 \psi_2 g_2\nonumber\\
   && + (2b_1 - b_3) (g_3)^2 - 2 \psi_3 g_3.
  \label{eq:fullvarF}
\end{eqnarray}
Minimizing this variational free energy yields the phase diagram for $\veta(s, T)$. Note that for the ordered, i.e., superconducting, side, we have to replace $\eta_i \mapsto \bar{\eta}_i + \delta \eta_i$ and additionally minimize with respect to $\bar{\eta}_i$ (see Appendix~\ref{app:free_en} for the resulting free energy).

Before we continue, we can gain some first insights from the self-consistency equations following from $\partial_{\psi_i} F_\Psi=0$,
\begin{eqnarray}
  \psi_0 &=& (6b_1+b_3) \int (d^3q) \frac{f_0(\q)}{f_0(\q)^2 - |\vec{f}(\q)|^2}\label{eq:psi0}\\
  \psi_1 &=& - (2b_2 + 2b_1 + b_3)\int (d^3q) \frac{f_1(\q)}{f_0(\q)^2 - |\vec{f}(\q)|^2}\label{eq:psi1}\\
  \psi_2 &=& (2b_2 - 2b_1  - b_3)\int (d^3q) \frac{f_2(\q)}{f_0(\q)^2 - |\vec{f}(\q)|^2}\label{eq:psi2}\\
  \psi_3 &=&  - (2b_1 -b_3) \int (d^3q) \frac{f_3(\q)}{f_0(\q)^2 - |\vec{f}(\q)|^2}\label{eq:psi3}.
\end{eqnarray}
The first equation describes the fluctuations in the order parameter, $\langle |\veta|^2\rangle$, which are non-zero for all temperatures. $\psi_1$ and $\psi_2$ describe fluctuations with a relative phase shift of 0 and $\pi/2$, ${\langle\eta_x^*\eta_y \pm \eta_y^*\eta_x\rangle}$, respectively, between $\eta_x$ and $\eta_y$. For $b_2>0$, only the fluctuations $\psi_2$ become non-zero and we thus set $\psi_1\equiv0$ in the following. Finally, $\psi_3$ are fluctuations that break the symmetry between $\eta_x$ and $\eta_y$, $\langle |\eta_x|^2 - |\eta_y|^2\rangle$. 

Note that for any of the fields $\psi_i$, $i=1,2,3$, the prefactor has to be positive in order to allow for a non-zero solution for zero strain in the normal state.
There is thus a region in parameter space, where the transition into the superconducting state is second order despite the additional $Z_2$ symmetry breaking (see dashed triangle in Fig.~\ref{fig:params}). It is currently not clear to us whether this is a real effect, or whether it is an artifact of the self-consistent harmonic approximation. In what follows, we will focus on parameters that are outside the dashed triangle in Fig.~\ref{fig:params}(b).
Outside this triangle, the superconducting transition is either first order, or it is preceded by a transition in which either $\psi_1$, $\psi_2$, or $\psi_3$ acquire a non-zero expectation value. These states corresponds to a time-reversal broken ($\psi_2$) and $C_4$ broken ($\psi_1$, $\psi_3$) phases, respectively.

\section{Results}
In order to minimize the variational free energy Eq.~\eqref{eq:fullvarF}, we perform the integrals numerically on a lattice, i.e., we replace $q_i^2 \mapsto 2 - 2\cos q_i$ and $q_xq_y \mapsto \sin q_x \sin q_y$. Note that this introduces a fixed ultra-violet cut-off. As a consequence, the absolute values of the parameters become important and not simply their ratios. We set the energy scale through $a_0 T_{\rm c}^{(0)} = 1$, and use the weak-coupling, circular Fermi surface  parameters $K_1 = 3K_2=3K_3=3K_4$, and a strong anisotropy $K_5=K_1/100$.~\cite{sigrist:2001}  

\begin{figure}[t]
  \centering
  \includegraphics{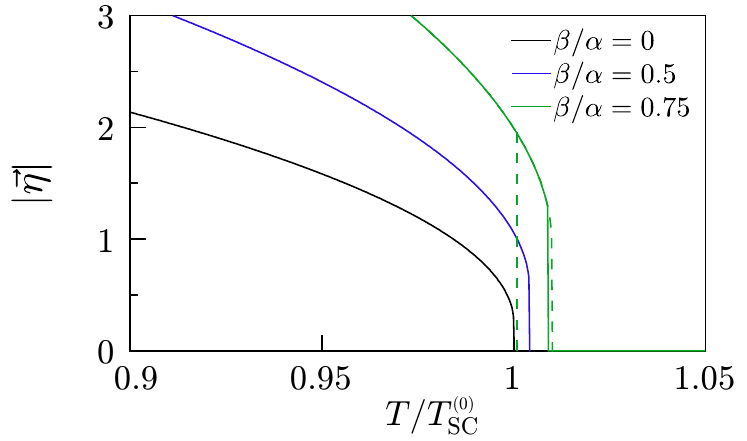}
  \caption{(Color online) Zero-strain order parameter for $\alpha=0.05$ and various TRS-breaking couplings $\beta/\alpha$. For $\beta/\alpha=0.75$, the points, where the metastable solutions disappear, are indicated as well (dashed lines). Note that for any finite $\beta/\alpha>0$, the transition becomes (weakly) first order.}
  \label{fig:s0}
\end{figure}

\subsection{Zero strain}
We start our discussion of the results for the case of zero strain, $s=0$. In this case, $\psi_1 \equiv \psi_3 \equiv 0$, and we are left with the two coupled equations
\begin{eqnarray}
  \tilde{a} &=& a_0(T-T_{\rm c}^{(0)}) + \alpha \int (d^3q) \frac{f_0(\q)}{f_0(\q)^2 - |\vec{f}(\q)|^2}\label{eq:atilde}\\
  \psi_2 &=& \beta \int (d^3q) \frac{f_2(\q)}{f_0(\q)^2 - |\vec{f}(\q)|^2}\label{eq:psi2b},
\end{eqnarray}
where, for simplicity, we have introduced $\tilde{a} = a_0(T-T_{\rm c}^{(0)}) + \psi_0$, $\alpha = 6b_1 + b_3$, and $\beta = 2b_2 - 2b_1 - b_3$. Note that this removes the explicit temperature dependence from Eq.~\eqref{eq:psi2b}.

Figure~\ref{fig:s0} shows the superconducting order parameter $|\veta|$ as a function of temperature for $\beta \geq 0$. For $\beta=0$, the system in general undergoes a second-order transition into a superconducting state when $\tilde{a}=0$.~\footnote{Note that for the fixed UV cut-off we have introduced by performing the integrals on a lattice, large $\alpha$ will drive the transition to be first-order. This limits the parameter range we can study numerically.} Since for $K_5>0$ the integral in Eq.~\eqref{eq:atilde} is bound by some constant $C$ from above for $\tilde{a}\rightarrow 0$, this transition occurs at a finite $T_{\rm SC}^{(0)} = T_{\rm c}^{(0)} - C/a_0$, with $T_{\rm c}^{(0)}$ the mean-field transition temperature.

\begin{figure}[t]
  \centering
  \includegraphics{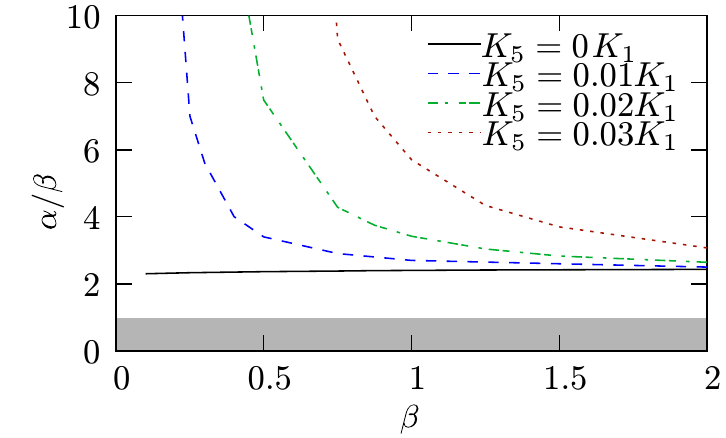}
  \caption{(Color online) The minimal ratio $\alpha/\beta$ needed in order to have a second-order transition into a non-superconducting TRS-breaking phase. The shaded region denotes $\alpha/\beta<1$, where the action Eq.~\eqref{eq:genaction} becomes unstable.}
  \label{fig:secondorder}
\end{figure}

For $\beta > 0$, however, TRS-breaking fluctuations in the order parameter develop above $T_{\rm SC}^{(0)}$ due to the divergence in the integral in Eq.~\eqref{eq:psi2b}, hence driving the superconducting transition. This can either lead to a combined first-order transition as in Fig.~\ref{fig:s0}, or there could be two consecutive transitions with a TRS-breaking, non-superconducting phase that precedes the superconducting phase. The situation is analogous to that of the magnetic ordering in the iron-based superconductors (see Refs.~[\onlinecite{Fang2008},\onlinecite{Xu2008},\onlinecite{fernandes:2012b}]), where the magnetic phase can be preceded by a nematic (C$_4$ breaking) phase. 
To analyze the possibility of a precursory time-reversal breaking phase, we follow the treatment in Ref.~[\onlinecite{fernandes:2012b}], first expressing $\tilde{a} = \tilde{a}(\psi_2)$ through Eq.~\eqref{eq:psi2b}. We then write Eq.~\eqref{eq:atilde} as a function of $\psi_2$ only, 
\begin{equation}
  a_0(T-T_{\rm c}^{(0)}) = \tilde{a}(\psi_2) - \alpha I(\psi_2).
  \label{eq:secondorder}
\end{equation}
For the right hand side of this equation, a maximum at $\psi_2=0$ leads to a first solution upon decreasing temperature at $\psi_2=0$, hence a second-order transition.
Since $\tilde{a}$ is a monotonically increasing function of $\psi_2$, the integral $I(\psi_2)$ needs to be an increasing function of $\psi_2$, too, for this to happen. 

Figure~\ref{fig:secondorder} shows the ratio $\alpha/\beta$ needed to have a second-order transition for various values of $K_5$. Note that the ratio $\alpha/\beta$ depends on the value of $\beta$. For the strictly two-dimensional case, a TRS-breaking phase forms at a finite $T_{\rm c}^{\rm TRS}$, while within the self-consistent harmonic approximation $T_{\rm SC}^{(0)}\equiv 0$ due to fluctuations. For finite $K_5$, the ratio diverges at a finite $\beta$; therefore, $\beta$ needs to exceed a critical value for a precursory TRS breaking phase to exist. Below this critical value, the transition into a TRS-breaking phase becomes first order, though there could still be a split transition.

\subsection{Full phase diagram}
\begin{figure}[t]
  \centering
  \includegraphics{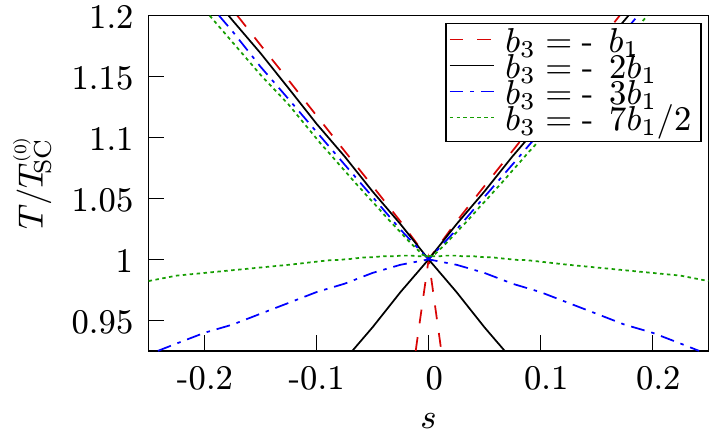}
  \caption{(Color online) Temperature-strain phase diagram for $b_2=5\times 10^{-4}b1$ and different values of $b_3$. The upper lines denote $T_{\rm c}^{(1)}$, where the system enters a $p_x$ ($p_y$)-wave state, and the lower lines $T_{\rm c}^{(2)}$, where it enters the chiral state. Note that the temperature is scaled with respect to the second-order transition temperature $T_{\rm SC}^{(0)}$.}
  \label{fig:b3}
\end{figure}

We now discuss how fluctuations change the temperature-strain mean-field phase diagram of Fig.~\ref{fig:params}(a). Here, we use for (numerical) simplicity the more isotropic parameter $K_5=K_1/3$. For finite strain, the shape of the phase diagram is mainly determined by the ratio $b_1/ b_3$. Figure~\ref{fig:b3} shows the phase diagram for a fixed transition temperature $T_{\rm SC}^{(0)}$ at zero strain, and $b_2=5\times 10^{-4}b_1$ for various $b_3$. As in the mean-field case, the main dependence on the quartic couplings is for the lower transition temperature $T_{\rm c}^{(2)}$, where the system enters the chiral phase. However, the fluctuations $\psi_3$ also shift $T_{\rm c}^{(1)}$, as they either increase ($b_3 > -2b_1$) or decrease ($b_3<-2b_1$) the effect of strain.

\begin{figure}[h]
  \centering
  \includegraphics{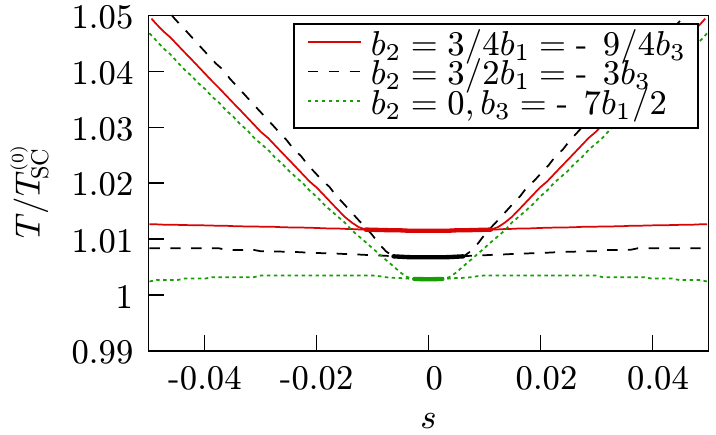}
  \caption{(Color online) Temperature-strain phase diagram for finite $b_2$ and $b_3$ showing no cusp at zero strain due to the first-order transition directly into a chiral state (thick lines) driven by the TRS-breaking fluctuations. Note that the temperature scale is with respect to the second-order transition temperature $T_{\rm SC}^{(0)}$.}
  \label{fig:b2b3}
\end{figure}

Figure~\ref{fig:b2b3} shows the phase diagram for smaller strains and $b_2\geq 0$. For $b_3<-2b_1$, there is a single, weakly first-order transition (see Fig.~\ref{fig:s0}) into a chiral superconducting phase driven by the TRS-breaking fluctuations. A finite strain is thus necessary to first condense into a $p_x$-- or $p_y$--wave phase, which smoothes the cusp at $s=0$. This effect is even stronger for $b_2>0$, see solid line in Fig.~\ref{fig:b2b3}.

\section{Discussion and Conclusions}
Before concluding, we comment on the relevance of our results to Sr$_2$RuO$_4$. While the exact values of the parameters in Eqs.~\eqref{eq:GLF2}-\eqref{eq:GLFgrad} depend on microscopic details and the form of the gap function, we can estimate the parameters based on a (naive) weak-coupling picture for the different bands.~\cite{agterberg:1997} For the two-dimensional $\gamma$ band,~\cite{ng:2000a, wang:2013, scaffidi:2014} assuming a circular Fermi surface yields $b_2=-b_3=2b_1/3$ and quasi-two-dimensional dispersions $K_1=3K_2=3K_3=3K_4 \gg K_5$.~\cite{mackenzie:2003} Thus, $2b_1 - b_2 - b_3<0$ and the mean-field transition remains unaffected by fluctuations [see Eq.~\eqref{eq:psi2}]. For superconductivity dominantly on the quasi-one-dimensional bands,~\cite{raghu:2010c, chung:2012} the situation might be more favorable. In this case, the two gaps are almost decoupled, i.e., $b_3 \approx -2b_1$, with small corrections of order $t'^2$, with $t'$ the energy scale of terms in the Hamiltonian connecting the two bands, such as inter-orbital hopping or spin-orbit coupling. The time-reversal-symmetry breaking term will be even smaller and $O(t'^4)$. Together with the much stronger (one-dimensional) fluctuations, this could indeed result in a TRS-breaking phase above the superconducting phase. Note, however, that there is no justification for such a weak-coupling description of Sr$_2$RuO$_4$ and interactions could dramatically change these parameters. While there is currently no evidence for either a weakly first-order or a precursory non-superconducting TRS breaking phase above $T_c$ in Sr$_2$RuO$_4$, the lack of a cusp in $T_c$ as a function of strain~\cite{hicks:2014} may motivate a more refined experimental examination of the issue.

To conclude, we have analyzed the effects of (thermal) fluctuations for a two-component $p$-wave superconductor. Within the self-consistent harmonic approximation, we have found that the fluctuations due to time-reversal-symmetry breaking can drive the transition at zero strain to become weakly first-order and have analyzed the possibility of a second-order transition into a non-superconducting TRS-breaking phase. Interestingly, we found a range of parameters, where there is a second-order transition into the superconducting state out of the normal state, despite the additional $Z_2$-symmetry breaking. Whether this is an artifact of our method could be analyzed within a renormalization-group scheme and is beyond the scope of the present work.
We have further discussed the relevance of our results for the case of Sr$_2$RuO$_4$. 
Finally, the physics of a preemptive TRS-breaking state might also be accessible in a recently proposed cold-atom setup.~\cite{buhler:2014}

\acknowledgements
We thank A.~Chubukov, M.~Sigrist, and R.~Fernandes for useful discussions. M.~H.~F. acknowledges support from the Swiss Society of Friends of the Weizmann Institute. E.~B. was supported by the ISF under grant 1291/12, by the BSF, and by a Minerva ARCHES grant.

\appendix
\section{Comparison to Hubbard-Stratonovich decoupling}
\label{app:hubb}
In order to perform a Hubbard-Stratonovich decoupling, we rewrite the quartic term in Eq.~\eqref{eq:GLF4} in terms of squares of quadratic terms. These terms correspond to the four fields $\psi_i$, $i=0,1,2,3$, namely
\begin{multline}
  f_4[\veta] = B_1(|\eta_x|^2 + |\eta_y|^2)^2
	+ B_2 (|\eta_x|^2 - |\eta_y|^2)^2\\
	+B_3(\eta_x^*\eta_y + \eta_x\eta_y^*)^2
	+B_4(\eta_x^*\eta_y - \eta_x\eta_y^*)^2
\end{multline}
It is important to note that these four terms are not linearly independent, namely the first minus the second term is the same as the third minus the fourth term. Therefore, this decoupling is not unique. However, for the calculation usually done after decoupling, namely the saddle-point approximation, a large-$N$ limit is assumed. Then, the terms have to be rewritten as $N$ component vectors, namely
\begin{multline}
  f_4[\veta] = B_1(|\vec{\eta}_x|^2 + |\vec{\eta}_y|^2)^2
	+ B_2 (|\vec{\eta}_x|^2 - |\vec{\eta}_y|^2)^2\\
	+B_3[\sum_i (\eta^i_x)^*\eta^i_y + \eta^i_x(\eta^i_y)^*]^2
	+B_4[\sum_i (\eta^i_x)^*\eta^i_y - \eta^i_x(\eta^i_y)^*]^2
\end{multline}
In the large-$N$ limit, the four terms are thus not linearly dependent anymore, such that a given decoupling in $N=1$ corresponds to a unique quartic term in large $N$.

\section{Free energy in the superconducting phase}
\label{app:free_en}
For completeness, we present here the variational free energy within the superconducting phase.
For this purpose, we start from Eqs.~\eqref{eq:GLF2}-\eqref{eq:GLFgrad} and expand $\eta_i=\bar{\eta}_i + \delta\eta_i$. For simplicity, we use $\bar{\eta}_x \in \mathbb{R}$ and $ i \bar{\eta}_y \in \mathbb{R}$. Then, we find for the elements of the inverse Green's function  
\begin{eqnarray}
  f_0(\q, \bar{\eta}) &=& f_0(\q) + \frac{6b_1+b_3}2(\bar{\eta}_x^2 + \bar{\eta}_y^2),\\
  f_1(\q, \bar{\eta}) &=& f_1(\q),\\
  f_2(\q, \bar{\eta}) &=& f_2(\q) + (2b_2-2b_1-b_3)\bar{\eta}_x\bar{\eta}_y,\\ 
  f_3(\q, \bar{\eta}) &=& f_3(\q) + \frac{2b_1-b_3}2(\bar{\eta}_x^2 - \bar{\eta}_y^2),
  \label{eq:fsSC}
\end{eqnarray}
with $f_i(\q)$ from Eqs.~\eqref{eq:psi0}-\eqref{eq:psi3}.
In addition, the free energy now reads
\begin{equation}
  F_{\Psi, \bar{\eta}}^{\rm SC}[\delta\veta] = F_{\Psi} + F[\bar{\eta}]
  \label{eq:FESC}
\end{equation}
with the variational free energy for the mean values
\begin{multline}
  F[\bar{\eta}] = a(\bar{\eta}_x^2 + \bar{\eta}_y^2) + s(\bar{\eta}_x^2 - \bar{\eta}_y^2)\\ + b_1(\bar{\eta}_x^2 + \bar{\eta}_y^2)^2 + (b_3 - b_2)\bar{\eta}_x^2\bar{\eta}_y^2.
  \label{eq:vareta}
\end{multline}
\bibliography{refs}

\begin{thebibliography}{23}%
\makeatletter
\providecommand \@ifxundefined [1]{%
 \@ifx{#1\undefined}
}%
\providecommand \@ifnum [1]{%
 \ifnum #1\expandafter \@firstoftwo
 \else \expandafter \@secondoftwo
 \fi
}%
\providecommand \@ifx [1]{%
 \ifx #1\expandafter \@firstoftwo
 \else \expandafter \@secondoftwo
 \fi
}%
\providecommand \natexlab [1]{#1}%
\providecommand \enquote  [1]{``#1''}%
\providecommand \bibnamefont  [1]{#1}%
\providecommand \bibfnamefont [1]{#1}%
\providecommand \citenamefont [1]{#1}%
\providecommand \href@noop [0]{\@secondoftwo}%
\providecommand \href [0]{\begingroup \@sanitize@url \@href}%
\providecommand \@href[1]{\@@startlink{#1}\@@href}%
\providecommand \@@href[1]{\endgroup#1\@@endlink}%
\providecommand \@sanitize@url [0]{\catcode `\\12\catcode `\$12\catcode
  `\&12\catcode `\#12\catcode `\^12\catcode `\_12\catcode `\%12\relax}%
\providecommand \@@startlink[1]{}%
\providecommand \@@endlink[0]{}%
\providecommand \url  [0]{\begingroup\@sanitize@url \@url }%
\providecommand \@url [1]{\endgroup\@href {#1}{\urlprefix }}%
\providecommand \urlprefix  [0]{URL }%
\providecommand \Eprint [0]{\href }%
\providecommand \doibase [0]{http://dx.doi.org/}%
\providecommand \selectlanguage [0]{\@gobble}%
\providecommand \bibinfo  [0]{\@secondoftwo}%
\providecommand \bibfield  [0]{\@secondoftwo}%
\providecommand \translation [1]{[#1]}%
\providecommand \BibitemOpen [0]{}%
\providecommand \bibitemStop [0]{}%
\providecommand \bibitemNoStop [0]{.\EOS\space}%
\providecommand \EOS [0]{\spacefactor3000\relax}%
\providecommand \BibitemShut  [1]{\csname bibitem#1\endcsname}%
\let\auto@bib@innerbib\@empty
%</preamble>
\bibitem [{\citenamefont {Sigrist}\ and\ \citenamefont
  {Ueda}(1991)}]{sigrist:1991}%
  \BibitemOpen
  \bibfield  {author} {\bibinfo {author} {\bibfnamefont {M.}~\bibnamefont
  {Sigrist}}\ and\ \bibinfo {author} {\bibfnamefont {K.}~\bibnamefont {Ueda}},\
  }\href@noop {} {\bibfield  {journal} {\bibinfo  {journal} {Rev. Mod. Phys.}\
  }\textbf {\bibinfo {volume} {63}},\ \bibinfo {pages} {239} (\bibinfo {year}
  {1991})}\BibitemShut {NoStop}%
\bibitem [{\citenamefont {Kallin}(2012)}]{kallin:2012}%
  \BibitemOpen
  \bibfield  {author} {\bibinfo {author} {\bibfnamefont {C.}~\bibnamefont
  {Kallin}},\ }\href@noop {} {\bibfield  {journal} {\bibinfo  {journal}
  {Reports on Progress in Physics}\ }\textbf {\bibinfo {volume} {75}},\
  \bibinfo {pages} {042501} (\bibinfo {year} {2012})}\BibitemShut {NoStop}%
\bibitem [{\citenamefont {Hicks}\ \emph {et~al.}(2014)\citenamefont {Hicks},
  \citenamefont {Brodsky}, \citenamefont {Yelland}, \citenamefont {Gibbs},
  \citenamefont {Bruin}, \citenamefont {Barber}, \citenamefont {Edkins},
  \citenamefont {Nishimura}, \citenamefont {Yonezawa}, \citenamefont {Maeno},\
  and\ \citenamefont {Mackenzie}}]{hicks:2014}%
  \BibitemOpen
  \bibfield  {author} {\bibinfo {author} {\bibfnamefont {C.~W.}\ \bibnamefont
  {Hicks}}, \bibinfo {author} {\bibfnamefont {D.~O.}\ \bibnamefont {Brodsky}},
  \bibinfo {author} {\bibfnamefont {E.~A.}\ \bibnamefont {Yelland}}, \bibinfo
  {author} {\bibfnamefont {A.~S.}\ \bibnamefont {Gibbs}}, \bibinfo {author}
  {\bibfnamefont {J.~A.~N.}\ \bibnamefont {Bruin}}, \bibinfo {author}
  {\bibfnamefont {M.~E.}\ \bibnamefont {Barber}}, \bibinfo {author}
  {\bibfnamefont {S.~D.}\ \bibnamefont {Edkins}}, \bibinfo {author}
  {\bibfnamefont {K.}~\bibnamefont {Nishimura}}, \bibinfo {author}
  {\bibfnamefont {S.}~\bibnamefont {Yonezawa}}, \bibinfo {author}
  {\bibfnamefont {Y.}~\bibnamefont {Maeno}}, \ and\ \bibinfo {author}
  {\bibfnamefont {A.~P.}\ \bibnamefont {Mackenzie}},\ }\href@noop {} {\bibfield
   {journal} {\bibinfo  {journal} {Science}\ }\textbf {\bibinfo {volume}
  {344}},\ \bibinfo {pages} {283} (\bibinfo {year} {2014})}\BibitemShut
  {NoStop}%
\bibitem [{\citenamefont {Nandkishore}(2012)}]{nandkishore:2012b}%
  \BibitemOpen
  \bibfield  {author} {\bibinfo {author} {\bibfnamefont {R.}~\bibnamefont
  {Nandkishore}},\ }\href@noop {} {\bibfield  {journal} {\bibinfo  {journal}
  {Phys. Rev. B}\ }\textbf {\bibinfo {volume} {86}},\ \bibinfo {pages} {045101}
  (\bibinfo {year} {2012})}\BibitemShut {NoStop}%
\bibitem [{\citenamefont {Bauer}\ \emph {et~al.}(2013)\citenamefont {Bauer},
  \citenamefont {Lutchyn}, \citenamefont {Hastings},\ and\ \citenamefont
  {Troyer}}]{bauer:2013}%
  \BibitemOpen
  \bibfield  {author} {\bibinfo {author} {\bibfnamefont {B.}~\bibnamefont
  {Bauer}}, \bibinfo {author} {\bibfnamefont {R.~M.}\ \bibnamefont {Lutchyn}},
  \bibinfo {author} {\bibfnamefont {M.~B.}\ \bibnamefont {Hastings}}, \ and\
  \bibinfo {author} {\bibfnamefont {M.}~\bibnamefont {Troyer}},\ }\href@noop {}
  {\bibfield  {journal} {\bibinfo  {journal} {Phys. Rev. B}\ }\textbf {\bibinfo
  {volume} {87}},\ \bibinfo {pages} {014503} (\bibinfo {year}
  {2013})}\BibitemShut {NoStop}%
\bibitem [{\citenamefont {Bojesen}\ \emph {et~al.}(2013)\citenamefont
  {Bojesen}, \citenamefont {Babaev},\ and\ \citenamefont
  {Sudb\o{}}}]{bojesen:2013}%
  \BibitemOpen
  \bibfield  {author} {\bibinfo {author} {\bibfnamefont {T.~A.}\ \bibnamefont
  {Bojesen}}, \bibinfo {author} {\bibfnamefont {E.}~\bibnamefont {Babaev}}, \
  and\ \bibinfo {author} {\bibfnamefont {A.}~\bibnamefont {Sudb\o{}}},\
  }\href@noop {} {\bibfield  {journal} {\bibinfo  {journal} {Phys. Rev. B}\
  }\textbf {\bibinfo {volume} {88}},\ \bibinfo {pages} {220511} (\bibinfo
  {year} {2013})}\BibitemShut {NoStop}%
\bibitem [{\citenamefont {Bojesen}\ \emph {et~al.}(2014)\citenamefont
  {Bojesen}, \citenamefont {Babaev},\ and\ \citenamefont
  {Sudb\o{}}}]{bojesen:2014}%
  \BibitemOpen
  \bibfield  {author} {\bibinfo {author} {\bibfnamefont {T.~A.}\ \bibnamefont
  {Bojesen}}, \bibinfo {author} {\bibfnamefont {E.}~\bibnamefont {Babaev}}, \
  and\ \bibinfo {author} {\bibfnamefont {A.}~\bibnamefont {Sudb\o{}}},\
  }\href@noop {} {\bibfield  {journal} {\bibinfo  {journal} {Phys. Rev. B}\
  }\textbf {\bibinfo {volume} {89}},\ \bibinfo {pages} {104509} (\bibinfo
  {year} {2014})}\BibitemShut {NoStop}%
\bibitem [{\citenamefont {Fernandes}\ \emph {et~al.}(2012)\citenamefont
  {Fernandes}, \citenamefont {Chubukov}, \citenamefont {Knolle}, \citenamefont
  {Eremin},\ and\ \citenamefont {Schmalian}}]{fernandes:2012b}%
  \BibitemOpen
  \bibfield  {author} {\bibinfo {author} {\bibfnamefont {R.~M.}\ \bibnamefont
  {Fernandes}}, \bibinfo {author} {\bibfnamefont {A.~V.}\ \bibnamefont
  {Chubukov}}, \bibinfo {author} {\bibfnamefont {J.}~\bibnamefont {Knolle}},
  \bibinfo {author} {\bibfnamefont {I.}~\bibnamefont {Eremin}}, \ and\ \bibinfo
  {author} {\bibfnamefont {J.}~\bibnamefont {Schmalian}},\ }\href@noop {}
  {\bibfield  {journal} {\bibinfo  {journal} {Phys. Rev. B}\ }\textbf {\bibinfo
  {volume} {85}},\ \bibinfo {pages} {024534} (\bibinfo {year}
  {2012})}\BibitemShut {NoStop}%
\bibitem [{\citenamefont {Nie}\ \emph {et~al.}(2014)\citenamefont {Nie},
  \citenamefont {Tarjus},\ and\ \citenamefont {Kivelson}}]{nie:2014}%
  \BibitemOpen
  \bibfield  {author} {\bibinfo {author} {\bibfnamefont {L.}~\bibnamefont
  {Nie}}, \bibinfo {author} {\bibfnamefont {G.}~\bibnamefont {Tarjus}}, \ and\
  \bibinfo {author} {\bibfnamefont {S.~A.}\ \bibnamefont {Kivelson}},\
  }\href@noop {} {\bibfield  {journal} {\bibinfo  {journal} {Proceedings of the
  National Academy of Sciences}\ }\textbf {\bibinfo {volume} {111}},\ \bibinfo
  {pages} {7980} (\bibinfo {year} {2014})}\BibitemShut {NoStop}%
\bibitem [{\citenamefont {Fang}\ \emph {et~al.}(2008)\citenamefont {Fang},
  \citenamefont {Yao}, \citenamefont {Tsai}, \citenamefont {Hu},\ and\
  \citenamefont {Kivelson}}]{Fang2008}%
  \BibitemOpen
  \bibfield  {author} {\bibinfo {author} {\bibfnamefont {C.}~\bibnamefont
  {Fang}}, \bibinfo {author} {\bibfnamefont {H.}~\bibnamefont {Yao}}, \bibinfo
  {author} {\bibfnamefont {W.-F.}\ \bibnamefont {Tsai}}, \bibinfo {author}
  {\bibfnamefont {J.}~\bibnamefont {Hu}}, \ and\ \bibinfo {author}
  {\bibfnamefont {S.~A.}\ \bibnamefont {Kivelson}},\ }\href@noop {} {\bibfield
  {journal} {\bibinfo  {journal} {Physical Review B}\ }\textbf {\bibinfo
  {volume} {77}},\ \bibinfo {pages} {224509} (\bibinfo {year}
  {2008})}\BibitemShut {NoStop}%
\bibitem [{\citenamefont {Xu}\ \emph {et~al.}(2008)\citenamefont {Xu},
  \citenamefont {M{\"u}ller},\ and\ \citenamefont {Sachdev}}]{Xu2008}%
  \BibitemOpen
  \bibfield  {author} {\bibinfo {author} {\bibfnamefont {C.}~\bibnamefont
  {Xu}}, \bibinfo {author} {\bibfnamefont {M.}~\bibnamefont {M{\"u}ller}}, \
  and\ \bibinfo {author} {\bibfnamefont {S.}~\bibnamefont {Sachdev}},\
  }\href@noop {} {\bibfield  {journal} {\bibinfo  {journal} {Physical Review
  B}\ }\textbf {\bibinfo {volume} {78}},\ \bibinfo {pages} {020501} (\bibinfo
  {year} {2008})}\BibitemShut {NoStop}%
\bibitem [{\citenamefont {Sigrist}(2002)}]{sigrist:2002}%
  \BibitemOpen
  \bibfield  {author} {\bibinfo {author} {\bibfnamefont {M.}~\bibnamefont
  {Sigrist}},\ }\href@noop {} {\bibfield  {journal} {\bibinfo  {journal}
  {Progress of Theoretical Physics}\ }\textbf {\bibinfo {volume} {107}},\
  \bibinfo {pages} {917} (\bibinfo {year} {2002})}\BibitemShut {NoStop}%
\bibitem [{\citenamefont {Giamarchi}(2003)}]{giamarchi:2003}%
  \BibitemOpen
  \bibfield  {author} {\bibinfo {author} {\bibfnamefont {T.}~\bibnamefont
  {Giamarchi}},\ }\href@noop {} {\emph {\bibinfo {title} {Quantum Physics in
  One Dimension}}},\ International Series of Monographs on Physics\ (\bibinfo
  {publisher} {Clarendon Press},\ \bibinfo {year} {2003})\BibitemShut {NoStop}%
\bibitem [{\citenamefont {Sigrist}\ and\ \citenamefont
  {Monien}(2001)}]{sigrist:2001}%
  \BibitemOpen
  \bibfield  {author} {\bibinfo {author} {\bibfnamefont {M.}~\bibnamefont
  {Sigrist}}\ and\ \bibinfo {author} {\bibfnamefont {H.}~\bibnamefont
  {Monien}},\ }\href@noop {} {\bibfield  {journal} {\bibinfo  {journal}
  {Journal of the Physical Society of Japan}\ }\textbf {\bibinfo {volume}
  {70}},\ \bibinfo {pages} {2409} (\bibinfo {year} {2001})}\BibitemShut
  {NoStop}%
\bibitem [{Note1()}]{Note1}%
  \BibitemOpen
  \bibinfo {note} {Note that for the fixed UV cut-off we have introduced by
  performing the integrals on a lattice, large $\alpha $ will drive the
  transition to be first-order. This limits the parameter range we can study
  numerically.}\BibitemShut {Stop}%
\bibitem [{\citenamefont {Agterberg}\ \emph {et~al.}(1997)\citenamefont
  {Agterberg}, \citenamefont {Rice},\ and\ \citenamefont
  {Sigrist}}]{agterberg:1997}%
  \BibitemOpen
  \bibfield  {author} {\bibinfo {author} {\bibfnamefont {D.~F.}\ \bibnamefont
  {Agterberg}}, \bibinfo {author} {\bibfnamefont {T.~M.}\ \bibnamefont {Rice}},
  \ and\ \bibinfo {author} {\bibfnamefont {M.}~\bibnamefont {Sigrist}},\
  }\href@noop {} {\bibfield  {journal} {\bibinfo  {journal} {Phys. Rev. Lett.}\
  }\textbf {\bibinfo {volume} {78}},\ \bibinfo {pages} {3374} (\bibinfo {year}
  {1997})}\BibitemShut {NoStop}%
\bibitem [{\citenamefont {Ng}\ and\ \citenamefont {Sigrist}(2000)}]{ng:2000a}%
  \BibitemOpen
  \bibfield  {author} {\bibinfo {author} {\bibfnamefont {K.~K.}\ \bibnamefont
  {Ng}}\ and\ \bibinfo {author} {\bibfnamefont {M.}~\bibnamefont {Sigrist}},\
  }\href@noop {} {\bibfield  {journal} {\bibinfo  {journal} {EPL (Europhysics
  Letters)}\ }\textbf {\bibinfo {volume} {49}},\ \bibinfo {pages} {473}
  (\bibinfo {year} {2000})}\BibitemShut {NoStop}%
\bibitem [{\citenamefont {Wang}\ \emph {et~al.}(2013)\citenamefont {Wang},
  \citenamefont {Platt}, \citenamefont {Yang}, \citenamefont {Honerkamp},
  \citenamefont {Zhang}, \citenamefont {Hanke}, \citenamefont {Rice},\ and\
  \citenamefont {Thomale}}]{wang:2013}%
  \BibitemOpen
  \bibfield  {author} {\bibinfo {author} {\bibfnamefont {Q.~H.}\ \bibnamefont
  {Wang}}, \bibinfo {author} {\bibfnamefont {C.}~\bibnamefont {Platt}},
  \bibinfo {author} {\bibfnamefont {Y.}~\bibnamefont {Yang}}, \bibinfo {author}
  {\bibfnamefont {C.}~\bibnamefont {Honerkamp}}, \bibinfo {author}
  {\bibfnamefont {F.~C.}\ \bibnamefont {Zhang}}, \bibinfo {author}
  {\bibfnamefont {W.}~\bibnamefont {Hanke}}, \bibinfo {author} {\bibfnamefont
  {T.~M.}\ \bibnamefont {Rice}}, \ and\ \bibinfo {author} {\bibfnamefont
  {R.}~\bibnamefont {Thomale}},\ }\href@noop {} {\bibfield  {journal} {\bibinfo
   {journal} {EPL (Europhysics Letters)}\ }\textbf {\bibinfo {volume} {104}},\
  \bibinfo {pages} {17013} (\bibinfo {year} {2013})}\BibitemShut {NoStop}%
\bibitem [{\citenamefont {Scaffidi}\ \emph {et~al.}(2014)\citenamefont
  {Scaffidi}, \citenamefont {Romers},\ and\ \citenamefont
  {Simon}}]{scaffidi:2014}%
  \BibitemOpen
  \bibfield  {author} {\bibinfo {author} {\bibfnamefont {T.}~\bibnamefont
  {Scaffidi}}, \bibinfo {author} {\bibfnamefont {J.~C.}\ \bibnamefont
  {Romers}}, \ and\ \bibinfo {author} {\bibfnamefont {S.~H.}\ \bibnamefont
  {Simon}},\ }\href@noop {} {\bibfield  {journal} {\bibinfo  {journal} {Phys.
  Rev. B}\ }\textbf {\bibinfo {volume} {89}},\ \bibinfo {pages} {220510}
  (\bibinfo {year} {2014})}\BibitemShut {NoStop}%
\bibitem [{\citenamefont {Mackenzie}\ and\ \citenamefont
  {Maeno}(2003)}]{mackenzie:2003}%
  \BibitemOpen
  \bibfield  {author} {\bibinfo {author} {\bibfnamefont {A.~P.}\ \bibnamefont
  {Mackenzie}}\ and\ \bibinfo {author} {\bibfnamefont {Y.}~\bibnamefont
  {Maeno}},\ }\href@noop {} {\bibfield  {journal} {\bibinfo  {journal} {Rev.
  Mod. Phys.}\ }\textbf {\bibinfo {volume} {75}},\ \bibinfo {pages} {657}
  (\bibinfo {year} {2003})}\BibitemShut {NoStop}%
\bibitem [{\citenamefont {Raghu}\ \emph {et~al.}(2010)\citenamefont {Raghu},
  \citenamefont {Kapitulnik},\ and\ \citenamefont {Kivelson}}]{raghu:2010c}%
  \BibitemOpen
  \bibfield  {author} {\bibinfo {author} {\bibfnamefont {S.}~\bibnamefont
  {Raghu}}, \bibinfo {author} {\bibfnamefont {A.}~\bibnamefont {Kapitulnik}}, \
  and\ \bibinfo {author} {\bibfnamefont {S.~A.}\ \bibnamefont {Kivelson}},\
  }\href@noop {} {\bibfield  {journal} {\bibinfo  {journal} {Phys. Rev. Lett.}\
  }\textbf {\bibinfo {volume} {105}},\ \bibinfo {pages} {136401} (\bibinfo
  {year} {2010})}\BibitemShut {NoStop}%
\bibitem [{\citenamefont {Chung}\ \emph {et~al.}(2012)\citenamefont {Chung},
  \citenamefont {Raghu}, \citenamefont {Kapitulnik},\ and\ \citenamefont
  {Kivelson}}]{chung:2012}%
  \BibitemOpen
  \bibfield  {author} {\bibinfo {author} {\bibfnamefont {S.~B.}\ \bibnamefont
  {Chung}}, \bibinfo {author} {\bibfnamefont {S.}~\bibnamefont {Raghu}},
  \bibinfo {author} {\bibfnamefont {A.}~\bibnamefont {Kapitulnik}}, \ and\
  \bibinfo {author} {\bibfnamefont {S.~A.}\ \bibnamefont {Kivelson}},\
  }\href@noop {} {\bibfield  {journal} {\bibinfo  {journal} {Phys. Rev. B}\
  }\textbf {\bibinfo {volume} {86}},\ \bibinfo {pages} {064525} (\bibinfo
  {year} {2012})}\BibitemShut {NoStop}%
\bibitem [{\citenamefont {B{\"u}hler}\ \emph {et~al.}(2014)\citenamefont
  {B{\"u}hler}, \citenamefont {Lang}, \citenamefont {Kraus}, \citenamefont
  {M{\"o}ller}, \citenamefont {Huber},\ and\ \citenamefont
  {B{\"u}chler}}]{buhler:2014}%
  \BibitemOpen
  \bibfield  {author} {\bibinfo {author} {\bibfnamefont {A.}~\bibnamefont
  {B{\"u}hler}}, \bibinfo {author} {\bibfnamefont {N.}~\bibnamefont {Lang}},
  \bibinfo {author} {\bibfnamefont {C.~V.}\ \bibnamefont {Kraus}}, \bibinfo
  {author} {\bibfnamefont {G.}~\bibnamefont {M{\"o}ller}}, \bibinfo {author}
  {\bibfnamefont {S.~D.}\ \bibnamefont {Huber}}, \ and\ \bibinfo {author}
  {\bibfnamefont {H.~P.}\ \bibnamefont {B{\"u}chler}},\ }\href@noop {}
  {\bibfield  {journal} {\bibinfo  {journal} {Nat Commun}\ }\textbf {\bibinfo
  {volume} {5}},\ \bibinfo {pages} {4504} (\bibinfo {year} {2014})}\BibitemShut
  {NoStop}%
\end{thebibliography}%
\end{document}